\documentclass[twocolumn,showpacs,prl,superscriptaddress]{revtex4}

\usepackage{amsmath}
\usepackage{tabularx}
\usepackage{graphicx}
\usepackage{color}
\usepackage{times}

\begin{document}

\title{Generalized black-box large deviation simulations:
High-precision work distributions for extreme
non-equilibrium processes in large systems}

\author{Alexander K. Hartmann}
\affiliation{Institut f\"ur Physik, Universit\"at Oldenburg, D-26111
	     Oldenburg, Germany}

\begin{abstract}
The distributions of work for strongly non-equilibrium 
 processes 
are studied using a very general form of a large-deviation approach,
 which allows one to study distributions of almost arbitrary quantities
of interest for equilibrium, 
non-equilibrium stationary and even non-stationary processes. 
The method is applied
to varying quickly  the external
field in a wide range 
$B=3\,\leftrightarrow 0$ for critical ($T=2.269$) two-dimensional 
Ising system of size $L\times L=128\times 128$. To obtain free energy
differences from the work distributions, they must be studied in ranges
where the probabilities are as small as $10^{-240}$, which
is not possible using direct simulation approaches. By comparison
with the exact free energies, which are available for this model
for the zero-field case,
one sees that the present approach
allows one to obtain the free energy with a very high relative precision
of $10^{-4}$. This works well also for non-zero field,
i.e., for a case where standard umbrella-sampling methods seem to be not so 
efficient to calculate free energies. Furthermore, for the present case 
it is verified that the resulting distributions of work 
for forward and backward process fulfill Crooks theorem with high precision.
Finally, the free energy for the Ising magnet as a function of the field
strength is obtained.
\end{abstract}


\maketitle


Studying non-equilibrium work processes \cite{evans2002,kurchan2007} 
has become a useful tool to extract information about physical systems.
Particular useful are the Jarzynski relation \cite{jarzynski1997} 
and the related
Crooks theorem \cite{crooks1998}, 
which allow one to extract equilibrium information
from non-equilibrium systems. Here, a system in
contact with a heat bath at temperature $T$ is prepared initially
in equilibrium, under the influence of some
external parameter $B=B_1$. Next it is driven out of equilibrium 
via quickly \cite{hendrix2001} varying $B=B_1\to B_2$, 
into another state, while performing some work $W$.
The Jarzynski relation relates the free  energy difference $\Delta F$
between the \emph{equilibrium} states at $B=B_1$ and $B=B_2$ to the
work $W$ performed during the non-equilibrium process $B=B_1\to B_2$:

\begin{equation}
\langle e^{-W/T} \rangle = e^{-\Delta F/T}\,,
\label{eq:jarzynski}
\end{equation}
where $\langle \ldots \rangle$ denotes the combined average over
the initial equilibrium distribution and all possible
process paths. In a similar way, the theorem of Crooks  
relates the distributions
of work $P(W)$ for the forward process and of the negated work 
$P_{\rm rev}(-W)$
for the reverse process $B=B_2\to B_1$ (where one starts in equilibrium at 
$B=B_2$) to the same $\Delta F$    via:

\begin{equation}
\frac{P(W)}{P_{\rm rev}(-W)} = e^{(W-\Delta F)/T}\,.
\label{eq:crooks}
\end{equation}
Hence, $P(W)$ and $P_{\rm rev}(-W)$ should intersect at $W^*=\Delta F$.
Unfortunately, the average of (\ref{eq:jarzynski}) and the region
where the two distributions intersect are \emph{often} 
dominated by exponentially
small probabilities, making finite-sampling errors particular strong
\cite{zuckerman2002}.
 Thus, the author of this work is only aware
of applications which exhibit either a small
number of degrees of freedom, or where the initial and final states
$B=B_1,B_2$ are very similar to each other. E.g., for the Ising
model in an external field $B$, work distributions have been obtained 
\cite{chatelain2006,hijar2007} for rather large
systems exhibiting $N=128^2$ spins, but successfully only in the paramagnetic
and in the ferromagnetic phases, 
where the work distribution could be approximated very well by a Gaussian.
At the critical point, it was only possible to 
sample the work distribution reliably
for very slow and small changes of the field, making the direct application
of the approach not successful.

As it is shown in this work, the work distributions can be
obtained very reliably via  large-deviation or 
importance-sampling techniques \cite{bucklew2004}, 
which are able to address large-deviation
regions \cite{touchette2009} of interest using bias functions.
Such techniques have been previously applied numerically to study
work distributions for small systems \cite{sun2003,ytreberg2004}.
If one targets not obtaining the full work distribution but 
is interested 
just in free energy differences,
it was suggested from results of simulations \cite{oberhofer2005}
and analytical studies \cite{lechner2007} of small model systems,
that applying work theorems
cannot compete with direct umbrella-sampling techniques
which explicitly obtain the distribution of the energy
over large ranges of the support. Thus, these results were
rather discouraging. 
Nevertheless, in the present work not only work distributions
are obtained over a large range of the support, but
it is also shown 
for a sample strongly non-equilibrium process
in a large system with a non-Gaussian
work distribution that a large-deviation non-equilibrium 
work-sampling approach turns out
to give very accurate results.

The algorithm presented in this work is a very general ``black-box'' type
approach which renders it applicable to study the distribution
of almost any quantity of interest for equilibrium, 
non-equilibrium stationary and even non-stationary processes. 
 The algorithm is here applied to work distributions of the Ising model in a 
non-zero field. In previous work 
the free energy could be obtained using umbrella-sampling approaches
for only rather small systems 
\cite{tsai2007}. This is in contrast
to the zero-field case, where indeed umbrella sampling is most efficient.
Here, the work distributions for the non-zero-field case are directly obtained
down to probabilities as small as $10^{-240}$
such that the Jarzynski relation (\ref{eq:jarzynski}) and 
Crooks theorem (\ref{eq:crooks}) can be
directly evaluated.  For this purpose,
the explicit biased sampling not only over the paths but also  over the initial
equilibrium distribution is included and simulations are performed for a 
large range of freely adjustable bias weights, allowing do obtain
the distributions over hundreds of decades in the probability.
The simulations are performed for large systems
and strongly non-equilibrium paths.

The ferromagnetic Ising model in a field $B\ge 0$ is studied here, given
by a set of $N$ Ising spins $s_i=\pm 1$ and described by
the Hamiltonian $H=-J\sum_{\langle i,j \rangle} s_i s_j - B \sum_i s_i$.
The first sum runs over all bonds connecting neighboring 
sites of a square lattice of size $L$, i.e., $N=L\times L$.
Periodic boundary conditions are applied in both directions.
The system is coupled to a heat bath at temperature $T=2.269$,
about the critical temperature for the 
ferromagnet-paramagnet phase transition.


Next, the numerical approaches are described. 
Processes were considered, where the system is started in equilibrium 
and within $n_{\rm iter}$ steps the field is changed.
For the \emph{forward} process it was 
increased from $B=0$ to $B=B_{\max}$, i.e., in each step by
$\Delta B= B_{\max}/n_{\rm iter}$. Thus, during each step $l$
the work $W_l=-\sum_i s_i \Delta B$ was performed,
the total work is $W=\sum_l W_l$.
After each field increment, one
sweep of a Monte Carlo (MC) simulation \cite{newman1999} with 
single-spin flip Metropolis dynamics
was performed. Hence, in each sweep, $N$ times a spin was randomly chosen
and a spin flip, exhibiting an energy change $\Delta H$,
was accepted with probability $\min\{1, \exp(-\Delta H/T) \}$.
The initial equilibrium configuration was obtained by starting
in a random configuration and performing 1000 steps of the Wolff cluster
algorithm \cite{wolff1989}, which should ensure equilibration since
the auto-correlation time for this algorithm is of the order 
of $\tau\approx 10$ at $T_c$ \cite{salas1996}. Also for $B_{\max}=3$ the
\emph{reverse} process was considered where the system
was started in equilibrium at $B=B_{\max}$ 
and the process $B=B_{\max}\to 0$ is performed in an analogous
way as for the forward process. 
Since the equilibrium configuration for this case is almost fully magnetized
at this large value of $B=B_{\max}$ (typically $0.002N$ spins are flipped),
the initial configurations were obtained by starting with all spins
up and performing one sweep of the single-spin flip dynamics prior
to the $B=B_{\max}\to 0$ process.

\begin{figure}
\begin{center}
\includegraphics[width=0.99\linewidth]{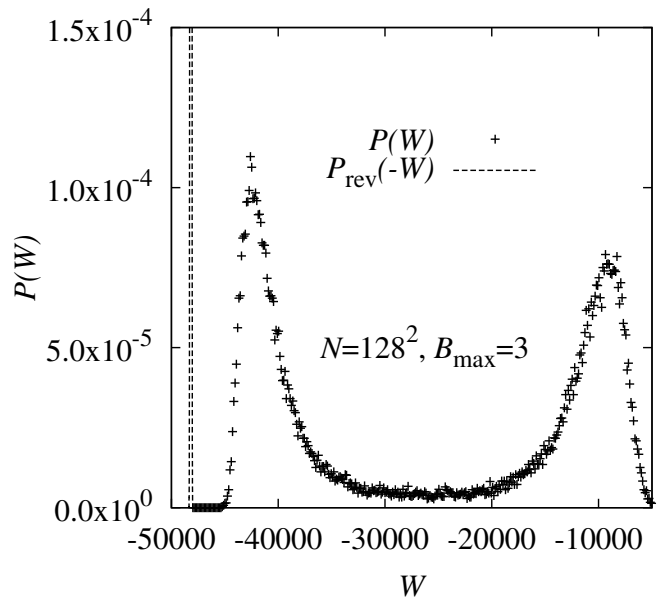}
\end{center}
\caption{Simple-sampling 
distribution $P(W)$ of work for an Ising system with $N=128^2$
spins for the forward $B=0\to 3$ and the mirrored distribution
$P_{\rm rev}(-W)$ for the reverse process $B=3\to 0$. 
\label{fig:PW_simple}}
\end{figure}

For the case $N=128^2$ and $10^6$ independent simulations, the histograms 
$P(W)$ of 
work for the forward and $P_{\rm rev}(-W)$ for the reverse process
are shown in Fig.\ \ref{fig:PW_simple}.
 According to the theorem of Crooks, the two histograms
should intersect at $W^*=\Delta F$. This appears to be somewhere between
$W=-50000$ and $W=-45000$, which on the first sight is only a small
interval compared the  support of the distribution $P(W)$
visible in Fig.\ \ref{fig:PW_simple}. Nevertheless, the probabilities
become so small, that it is impossible
to see the intersection using any feasible number of standard simulations runs.
Actually, as it is shown below, the crossing appears where $P(W)=
P_{\rm rev}(-W)\approx 10^{-57}$. Hence, numerical large deviations 
techniques have to be used, to address this region.

The algorithm presented here is different compared to 
well-known  algorithms as, e.g., the ``cloning'' approach 
\cite{giardina2006,lecomte2007,giardina2011}, and
consists of a second MC-simulation level.
Each
configuration is represented by a vector $\xi=(\xi_1,\xi_2,\ldots,\xi_M)$
of suitable length $M$ (see below). The basic idea is that the entries of 
$\xi$ are random variables uniformly distributed in $[0,1]$, which
are used to feed the random process under investigation. Hence, e.g.,
when performing the work distributions, the random decisions are not
based on numbers drawn from random number generators, but, in a
defined manner, on the entries of $\xi$. Here, for the forward process,
the first $n_{\rm Wolff}(2N+1)$ entries are used to feed $n_{\rm Wolff}=10$
iterations of the Wolff algorithm, starting from a precomputed
(1000 Wolff iterations) equilibrium configuration $s^{(0)}$. 
This allows to sample the equilibrium distribution prior to the work
process. Each Wolff
iteration consist of choosing one seed spin (consuming one entry of $\xi$)
plus a cluster growth, where possibly for each of the $2N$ bonds
it has to be decided randomly whether it is ``activated'' or not, for
details of the Wolff algorithm see Ref. \cite{wolff1989}. Note that
each bond is assigned a specific entry of $\xi$ (for each Wolff iteration,
respectively), independent of whether the bond is tried
to be activated or not. Next, the $n_{\rm iter}$ work sweeps
are performed, consisting of $n_{\rm iter}-1$ single-spin-flip
MC sweeps (the last sweep after the final field increment can be omitted), 
where for each sweep $2N$ entries of $\xi$ are consumed,
one for randomly selecting a spin, and one for the Metropolis criterion
(also if $\Delta H <0$). Hence, for one full process $M=n_{\rm Wolff}(2N+1)+
(n_{\rm iter}-1)2N$ entries, corresponding to random numbers, are used.
For the reverse process, where no Wolff algorithm is used,
$M=2N+(n_{\rm iter}-1)2N=n_{\rm iter}2N$.
Each process fed by $\xi$ results in a work $W(\xi)$. Now, the second-level MC
consist of changing a small number (here 400) of the entries of $\xi$, 
each drawn again uniformly from $[0,1]$, leading
to a ``trial configuration'' $\xi'$ with corresponding work $W(\xi')$.
The new configuration is accepted with the Metropolis probability
$\min\{1, \exp(-(W(\xi')-W(\xi))/T_{\rm MC}) \}$, otherwise $\xi$ is
kept for the next second-level MC step. Thus, the observed distribution
of work will exhibt a Boltzmann bias
$P_{T_{\rm MC}}(W)\sim P(W)e^{-W/T_{\rm MC}}$. Note that $T_{\rm MC}$
is a freely adjustable parameter, different from the temperature $T$
of the Ising system, which allows to center the observed distribution in
different regions.
Therefore, by
performing the simulation for different values of $T_{\rm MC}$ such that
the resulting distributions $P_{T_{\rm MC}}(W)$ overlap, one can reconstruct
the desired distribution $P(W)$ via reweighting and gluing the distributions
together \cite{align2002}. 
Note that instead of using a Boltzmann bias for the observed
work, one could also use an umbrella sampling on the $\xi$ vectors 
to obtain
$P(W)$ directly. However, this was tried during the present work 
extensively, but the Boltzmann bias was more efficient.


Seen in an abstract manner, the approach is based on
a configuration vector $\xi$, an evaluation function $\tilde H(\xi)$,
 a simple dynamics changing
 the configuration vector to create trial vectors $\xi'$ 
and a Metropolis criterion (involving ``temperature'' $T_{\rm MC}$), 
which accepts $\xi$ based on $\tilde H(\xi)$ and $\tilde H(\xi')$. 
All problem-specific details are included in the function
$\tilde H$.
Hence, if $\tilde H$ was the energy of a spin system, 
then the second-level
MC would be a standard MC simulation. But $\tilde H$ can
represent almost any process, like for the present application
where it states the work $W$ arising from an equilibrium sampling
followed by
a strongly non-equilibrium work process.
Using this
general view on the large-deviation simulation, it is clear that
the approach can be applied basically to any equilibrium, non-equilibrium
stationary and even non-stationary process which transforms
a vector of a random numbers into a measurable result, independent
of how involved this transformation is. Hence, the approach should
be applicable to \emph{any}  random 
process which can be simulated on a computer.


\begin{figure}
\begin{center}
\includegraphics[width=0.99\linewidth]{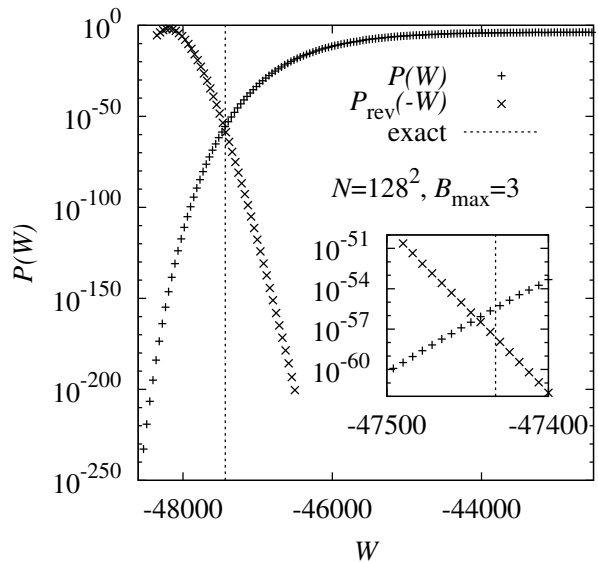}
\end{center}
\caption{Distribution of works for the forward and mirrored distribution
for the reverse process for $B=0\leftrightarrow 3$, $L=128$.
 According to the theorem of Crooks, these distributions
are expected to intersect at $W^*=\Delta F$ signifying the 
free-energy difference $\Delta F$. The exact free-energy difference is
indicated by the vertical line.
The inset shows a blow up of the intersection region.
\label{fig:PW_full}}
\end{figure}

For the simulations, a number between 4 ($B_{\max}=0.25$) 
and 37 ($B_{\max}=3$)  of temperatures 
$T_{\rm MC}\in [1.5, 200]$  were considered. 
The spacing $\Delta T_{\rm MC}$ 
between the MC temperatures ranged from 0.1 for low values
of $T_{\rm MC}$ up to 100 for larges values. For each value of $T_{\rm MC}$, 
$10^6$ MC trials were performed, taking about 6 hours on a core 
of standard 2.66 GHz Intel Westmere processor, i.e., just $37\times 6=222$
core hours for the strongest field  $B_{\max}=3$.

Concerning the analysis of the simulation leading to the full
work distribution,
the intersection region of 
the distributions of work for the forward and reverse processes
are shown in Fig. \ref{fig:PW_full} for $B_{\rm max}=3$.
The two distributions $P(W)$ and $P_{\rm rev}(-W)$ intersect
at $W^*\approx -47443$. For comparison, also the exact free energy
difference was obtained. For the zero-field case, the exact free
energy $F_0$ is known analytically \cite{ferdinand1969}
for finite-size systems.
For the case $B>0$, if $B$
is large, the system is almost fully magnetized, except for a few
single-spin excitations. Hence, the free energy is given via
$F_B = -T \log \left ( e^{(2+B)N/T}\left[1+ \sum_{k=1}^N {N \choose k} 
\exp^{-(8+2B)k/T}  \right]\right)$, where only few terms of the sum
around the typical number $k$ of excited spins (about $0.002N$ for $B=3$)
contribute significantly.  For $L=128,B=3.0$, 
this results in $\Delta F=F_B-F_0\approx-47433$. Thus,
the relative deviation of the estimated from the exact free energy
difference is only $(\Delta F - W^*)/\Delta F=0.0002$.

\begin{figure}
\begin{center}
\includegraphics[width=0.99\linewidth]{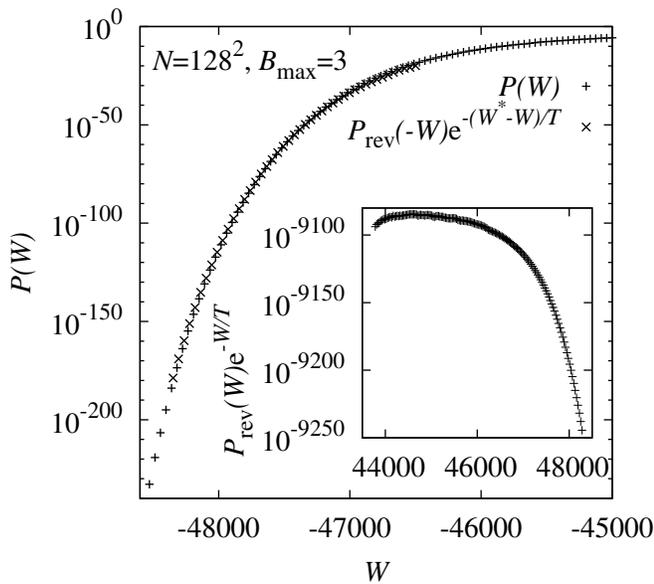}
\end{center}
\caption{Distribution of works for the forward and 
rescaled-mirrored distribution for the reverse process. 
The inset exhibits the integrand $P_{\rm rev}(W)\exp(-W/T)$ used to obtain
$\langle \exp(-W/T) \rangle$ from the reverse process $B=3\to 0$. 
\label{fig:PW_scale}}
\end{figure}

To verify whether the data fulfills Crooks theorem (\ref{eq:crooks}),
the histogram for the reverse process was rescaled accordingly, see
Fig.\ \ref{fig:PW_scale}. This is
confirmed by the data with high precision. Note that
testing Crooks theorem may also conveniently serve as
a check that the second-level MC simulations are equilibrated.

\begin{figure}
\begin{center}
\includegraphics[width=0.99\linewidth]{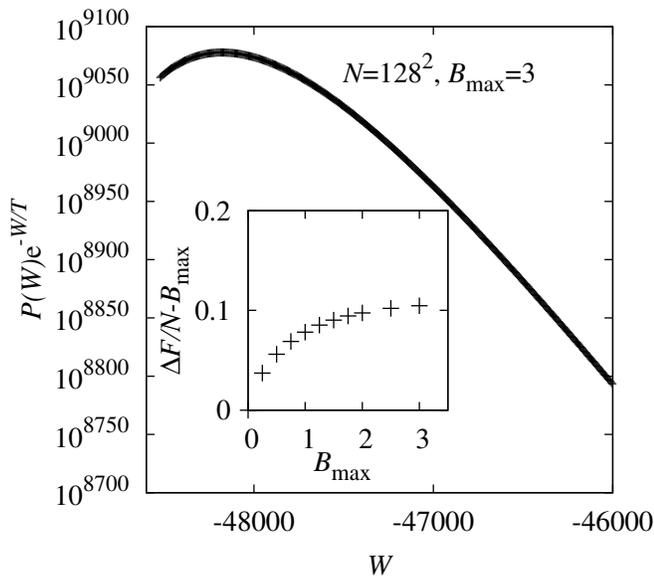}
\end{center}
\caption{Integrand $P(W)\exp(-W/T)$ used to obtain
$\langle \exp(-W/T) \rangle$. Inset: Resulting free energy 
difference $\Delta F$ per spin $N$ (minus $B_{\max}$)
as a function of the final magnetic field $B_{\max}$ for the forward process.
\label{fig:PW_integrand}}
\end{figure}

To obtain $\Delta F$ using the Jarzynski relation (\ref{eq:jarzynski}), 
the integral
$\langle  \exp(-W/T) \rangle = \int \exp(-W/T)\,P(W)\,dW$ has to
be evaluated, resulting in $\Delta F \approx -47438$, which has
a relative deviation 0.0001 from the exact result.
 The integrand is shown in Fig.\ \ref{fig:PW_integrand}.
Note that only the region close to the peak around $W=-48100$ contributes
significantly to the integral, which deviates
much from the point where $P(W)$ and $P_{\rm rev}(-W)$
intersect.
 Nevertheless, one has to obtain the
full distribution in a region ranging from its peak value at $W=-42000$
down to $W=-48200$ to get $P(W)$ right. 
For the reverse process, the evaluation 
(see inset of Fig.\
\ref{fig:PW_scale}) results in $-\Delta F= 47450$, which deviates
by a factor of $0.0004$ from the exact value. Note that one has
to obtain $P_{\rm rev}(W)$ over an even broader support, which is probably
the reason for the somehow smaller accuracy compared to the forward process.

The forward process was performed for different values of $B_{\rm max}$,
see inset of Fig.\ \ref{fig:PW_integrand}.
 The amount by which $\Delta F/N$ is larger than $B_{\rm max}$
describes the entropy loss due to the alignment of the spins to the field.


To summarize, a biased sampling approach is introduced, which 
is based only on a Markov-chain evolution of a vector of
entries from the interval $[0,1]$,
seen as an input vector of random numbers to an arbitrary stochastic process,
 which can be treated as a black box
within the approach.

Here, high-precision determination 
of work
distributions for Ising magnets in a field 
were performed, for large systems and strongly non-equilibrium
processes, hence
for cases where traditional direct approaches for measuring work
distributions completely fail. In the past only close-to
equilibrium processes could be studied with small accuracy 
\cite{chatelain2006,hijar2007}.
Still, the path sampling applied here
is very general, since no
details of the path construction have to be known to efficiently
sample the corresponding work distribution down to probabilities
as small as $10^{-240}$.
This contrasts the approach with previous  problem-specific 
yet quite successful techniques, like the
``shooting approach'' \cite{bolhuis2002} or ``cloning''
\cite{giardina2006,lecomte2007,giardina2011}.

The work distributions are used to extract free energy
differences for Ising systems in a field
using the Jarzynski relation \cite{jarzynski1997} and the theorem
of Crooks \cite{crooks1998}.
Note that for determining the free energy of the Ising system 
\emph{without
a field}, very good other approaches exist. Using the 
convenient Wang-Landau umbrella sampling 
\cite{wang2001},
 the free energy was determined very
accurately for systems of size $N=256^2$.  
Based on measuring the large-deviation properties of the number
of components for Fortuin-Kasteleyn clusters
 even systems of size
$N=1000^2$ could be treated \cite{partition2005}. Recently it
was claimed 
\cite{oberhofer2005} that in general umbrella-sampling approaches 
should be superior or at least equally efficient as
applying large-deviation techniques to work distributions to
measure free-energy differences. Nevertheless, for the present
study of an Ising systems \emph{in a field},
using umbrella-sampling 
only sizes of $N=42^2$ could be studied so far \cite{tsai2007},
about ten times smaller than  the sizes addressed in the present study.
Hence, for certain systems, e.g., 
for an Ising magnet in a field, 
 almost the full work distribution can be determined using the present
approach. But even aiming only at 
determining free energy differences, the
present very general approach might be superior to highly-evolved 
existing techniques.
The observed failure of Refs.\ \cite{oberhofer2005,lechner2007} might be
due to the fact that there only very small systems could be studied.
Also it could be due to  
the specific ``shooting'' algorithm \cite{bolhuis2002} which might
not be most suitable for fully random processes.
 Finally, for some past studies also single \cite{ytreberg2004}
or very many \cite{sun2003}
specific bias
functions where used, while here a Boltzmann reweighting for
few selected weights was performed, which allows to
address different regions of interest independently.

Due to the black-box structure of the algorithm presented here, 
it allows to study equilibrium,
non-equilibrium stationary and even non-stationary systems. Hence,
for future work, many applications of this algorithm
can be anticipated.

\begin{acknowledgments} The author is grateful to
 Andreas Engel for useful discussions and critically reading the manuscript.
The authors thanks Oliver
Melchert for also critically reading the manuscript.
The simulations were performed on the HERO cluster of the University
of Oldenburg jointly funded by the DFG (INST 184/108-1 FUGG) and the 
ministry of Science and Culture (MWK) of the Lower Saxony State.
\end{acknowledgments}

\bibliography{alex_refs}

\end{document}